\documentclass[useAMS]{mn2e}
\usepackage{lscape}
\usepackage{rotating}

\def\cm3{\hbox{cm$^{-3}$}}

\newcommand{\db}{D{$_{\rm break}$}}		
\newcommand{\nmin}{N{$_{\rm min}$}}		
\newcommand{\mbar}{$\overline{\rm m}$}	
\newcommand{\sbar}{$\overline{\rm s}$}

\newcommand{\mcl}{M$_{\rm cl}$}	
\newcommand{\rcl}{R$_{\rm cl}$}	
\newcommand{\vcl}{$\sigma_{\rm cl}$}

\input psfig.sty
\input epsf.sty
\usepackage{graphicx}
\title[The evolution of structure in the LMC]
{The spatial evolution of stellar structures in the LMC}
\author[Bastian et al.]{Nate Bastian$^{1,2}$, Mark Gieles$^3$, Barbara Ercolano$^{1,4}$, Rob Gutermuth$^4$ \\
$^1$ Institute of Astronomy, University of Cambridge, Madingley Road, Cambridge,
 CB3 0HA, UK\\
 $^2$ Department of Physics and Astronomy, University College London, Gower Street, London, WC1E 6BT\\
$^3$ European Southern Observatory, Casilla 19001, Santiago 19, Chile  \\
$^4$ Harvard-Smithsonian Center for Astrophysics, 60 Garden Street, Cambridge, MA 02138, USA \\
}
\date{Accepted. Received; in original form}
\pagerange{\pageref{firstpage}--\pageref{lastpage}}
\pubyear{2008}
\begin{document}
\maketitle
\label{firstpage}
\begin{abstract}
We present an analysis of the spatial distribution of various stellar populations within the Large Magellanic Cloud.  We combine mid-infrared selected young stellar objects, optically selected samples with mean ages between $\sim9$ and $\sim1000$~Myr, and existing stellar cluster catalogues to investigate how stellar structures form and evolve within the LMC.  For the analysis we use Fractured Minimum Spanning Trees, the statistical $Q$ parameter, and the two-point correlation function.   Restricting our analysis to young massive (OB) stars we confirm our results obtained for M33, namely that the luminosity function of the groups is well described by a power-law with index $-2$, and that there is no characteristic length-scale of star-forming regions.  We find that stars in the LMC are born with a large amount of substructure, consistent with a 2D fractal distribution with dimension $\sim1.8$ and evolve towards a uniform distribution on a timescale of $\sim175$~Myr.    This is comparable to the crossing time of the galaxy and we suggest that stellar structure, regardless of spatial scale, will be eliminated in a crossing time. This may explain the smooth distribution of stars in massive/dense young clusters in the Galaxy, while other, less massive, clusters still display large amounts of structure at similar ages.  By comparing the stellar and star cluster distributions and evolving timescales, we show that infant mortality of clusters (or 'popping clusters') have a negligible influence on galactic structure.  Finally, we quantify the influence of the elongation, differential extinction, and contamination of a population on the measured $Q$ value.

\end{abstract}
\section{Introduction}\label{intro}

This study is part a series to quantify the degree and evolution of substructure within galaxies.  Here we introduce the methods and use the LMC as a test case.   The distribution of young stars in the LMC was first studied with objective algorithms by Feitzinger \& Braunsfurth (1984), who found that the distribution was not random but had distinct patterns on all of the spatial scales probed (60~pc to 8~kpc).    Other studies have come to similar conclusions, that young stars are distributed hierarchically within galaxies (e.g. Efremov~1984, Battinelli, Efremov, \& Magnier~1996, Elmegreen \& Salzer~1999, Elmegreen \& Elmegreen~2003, Elmegreen et al.~2006, Bastian et al.~2005, 2007, S{\'a}nchez et al.~2007).  This pattern observed in the young stellar distribution is thought to originate from the distribution and fragmentation of gas within galaxies (e.g. Elmegreen \& Efremov~1996; Elmegreen \& Falgarone~1996).


It has been noted that many nearby star forming clusters also have  
hierarchical structure seemingly dictated by the structure of the  
dense gas of natal molecular clouds (e.g. Lada \& Lada 2003).  Using  
statistical techniques, Gutermuth et al. (2005) studied three  
Galactic clusters of varying degrees of embeddedness and demonstrated that the  
least embedded cluster was also the least dense and the least  
substructured of the three. That result hinted at the idea that the  
youngest clusters are substructured, but that dynamical interactions  
and ejection of the structured gas contributes to the evolution and  
eventual erasure of that substructure in approximately the cluster  
formation timescale of a few Myr (Palla \& Stahler 2000). By having an  
accurate model of the spatial evolution of stellar structures we can  
approach a number of fundamental questions about star-formation,  
including what is the percentage of stars born in "clusters" and  
whether this depends on environmental conditions. Using automated  
algorithms on infrared Spitzer surveys of star-forming sites within  
the Galaxy (e.g. Allen et al. 2007), such constraints are now becoming  
possible.

In addition to constraining models of star-formation on galaxy and cluster scales, the spatial distribution of stars can have important implications on a variety of other astrophysical problems.  For example, the mixture of gas and stars can affect the temperature structure within H{\sc ii} regions which in turn can influence metallicity determinations based on emission line spectra (Ercolano, Bastian, Stasinska~2007).  Thus, a small dependence on clustering properties of young stars on galactocentric distance (i.e. if there is a slight preference to form more concentrated stellar groups in the central parts of galaxies compared to the outer regions as may be expected if gas pressure plays a dominant role in cluster formation) could artificially enhance the metallicity gradient observed in spiral galaxies.


Previous works on stellar structures have mainly concentrated on the youngest stellar objects (OB stars), while the evolution of these objects have remained largely unstudied.  Efremov~(1995) studied the distribution of Cepheids, whose ages are significantly higher than OB stars,  and found evidence of huge stellar complexes which he suggested may be made up of dissolved aged associations.  Zaritsky et al.~(2004) showed the stark contrast between the spatial distribution of young and old stellar populations in the LMC using two colour-magnitude cuts through the  Magellanic Clouds Photometric Survey (MCPS).  

The goal of this paper is to quantify the timescale over which stellar structures within the LMC evolve.  To this end, we use two methods to estimate the evolution and analyse each method's strengths and weaknesses.  Our primary dataset comes from the MCPS where we make cuts in colour and magnitude space in order to select differing stellar populations (quantifying the age through comparison with stellar isochrones with a given initial mass function) which we compliment with catalogues of stellar clusters (Hunter et al.~2003) and embedded young stellar objects (Whitney et al.~2008).

The paper is organised in the following way.  In \S~\ref{sec:obs} we introduce the observations and the methods used, while in \S~\ref{sec:size} we compare the properties (size and luminosity distributions) of the young star-forming groups between the LMC and M33.  In \S~\ref{sec:subsamples} we extract twelve samples of stars of varying mean ages in order to compare their clustering properties, which is carried out in \S~\ref{sec:spatial}.  In \S~\ref{sec:comparison} we discuss the implications of our results and present a series of simple models to aid in the interpretation, and in \S~\ref{sec:conclusions} we present our conclusions.

\section{Observations and Method}\label{sec:obs}

Our primary set of data comes from the Magellanic Cloud Photometry Survey of Zaritsky et al.~(2004).  This dataset consists of U, B, V, and I band imaging of essentially the entire optical extent of Large Magellanic Cloud.  The original list consists of over 24 million sources, far too many for our algorithm, so we restrict our study to various cuts in colour and magnitude space.

The method that we employ to first study the distribution of star-forming sites is the same as was presented in Bastian et al. (2007) for the study of the spatial distribution of OB stars in M33.  We refer the reader to that work for details and only provide a summary of the method here.  Using the photometry of individual stars we make colour and magnitude cuts in order to select young stars (see \S~\ref{sec:size} for details of the cuts).  We then construct a minimum spanning tree (MST) of the spatial positions of stars which pass our criteria\footnote{An MST is formed by connecting all points (spatial positions in this case)  in order to form a unified network, such that the total length (i.e. sum)  of all of the connections, known as 'edges' or 'branches', is minimized, and no closed loops are formed.}. The tree is then fractured (FMST) by applying a breaking distance, \db, in which all edges of the tree, longer than \db, are broken (i.e. removed), creating independent groups.  In order to avoid a selection bias which has hampered previous studies, we apply multiple breaking distances (20 in all) from pc to kpc scales.  Throughout this paper, unless otherwise noted, we have used a value of five for \nmin, which is the minimum number of sources used to define a group.

In \S~\ref{sec:subsamples} we extend our study by not only looking at the youngest stars, but also the distribution of stars of various ages.

Throughout this study we have assumed a distance to the LMC of 50 kpc (e.g. Marconi \& Clementini~2005).  For comparison with models (both fractal and uniform distributions) we will assume that the LMC is largely 2D - i.e. that it is a disk-like face on galaxy.  This assumption will be justified in \S~\ref{sec:q}.

\section{Size and luminosity distributions}
\label{sec:size}

As a first step towards understanding the spatial distribution of star-forming regions in the LMC we perform a similar study as was carried out on M33 (Bastian et al.~2007).  We have used cuts in colour and magnitude space (namely $B-V < 0.5$ and $M_{V} < -4.5$ - uncorrected for extinction) in order to select young (massive) stars, resulting in 5093 stars which pass these criteria.   The sample is then analysed using the Fractured Minimum Spanning Tree method, selecting 20 breaking distances between the effective resolution limit of 1.5~pc and 1~kpc (with equal steps in logarithmic space of 0.1485).  We refer the reader to Bastian et al. (2007) for details on the method.  This method produces a catalogue of groups and their properties, namely size and luminosity, for each breaking distance applied.  We also combined these catalogues into a 'total sample' from which we removed groups found by multiple breaking radii (i.e. duplicate groups that had the exact same position and radius) only retaining the groups found by the smallest breaking radius.  We note that in the individual catalogues we did not remove duplicates (i.e. groups that were found by other \db), as each breaking distance was treated seperately.

\subsection{Size distribution}

We first look at the size distribution of the groups found.  In Bastian et al.~(2007) we found that for M33 there was not a characteristic size for star-forming regions, contrary to what is often quoted in the literature.  We showed that previous claims of a characteristic size of $\sim100$~pc for OB-associations (e.g. Bresolin et al.~1998) was due to the selection of a single breaking distance or scale in which to carry out the analysis.

\begin{figure}
\begin{centering}
\hspace{-0.5cm}
\includegraphics[width=8.5cm]{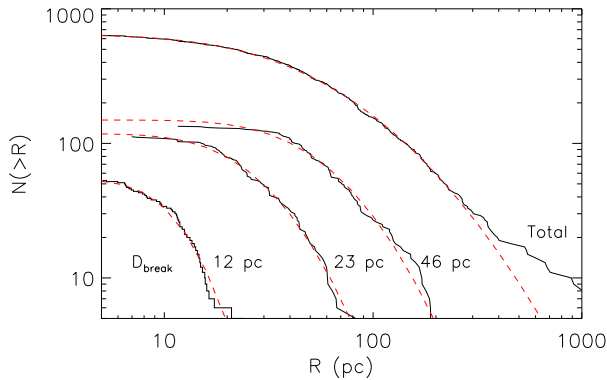}
\caption{The size (radius) distribution of the groups for three different breaking radii as well as for the total sample.  Each distribution is well fit by a log-normal distribution whose peak depends on the scale probed.  The dashed (red) lines show the best fitting log-normal function to each distribution.} 
\label{fig:sizefunc}
\end{centering}
\end{figure}

We carried out a similar analysis for the LMC, and come to similar conclusions.  Figure~\ref{fig:sizefunc} shows the cumulative size distribution of groups found with various breaking distances and for the total sample.    The dashed (red) lines show the best fitting log-normal function to each distribution, which provides a good fit for each breaking distance.   The total sample displays an excess tail at large radii ($> 400$~pc) relative to the best fitting analytical fit, however it is well fit up until this point.  This break is similar, although at slightly smaller scales (400~pc vs. 1~kpc) to that seen in the autocorrelation function of young stars in the LMC (Odekon~2008, her Fig.~3).    It is also similar to the scale where a break in the H{\sc I} distribution is observed, $180-290$~pc (Padoan et al.~2001; Elmegreen et al.~2001; Kim \& Park~2007).  This feature, seen in the H{\sc I} and in the young stars, has been suggested to relate to the scale height of the galaxy, where structures begin to change from largely 3D to 2D objects.  We note that Odekon~(2008) also found a break in the autocorrelation function for young stars in M33, whereas no similar scale was identified in the study of Bastian et al.~(2007).


We confirm the results found for M33, that the turn-over in the size-distribution depends on the breaking distance used in an almost one-to-one way, suggesting that previous reports of a characteristic size in star-forming regions were due to the selection of a single breaking distance.  Hence, we confirm that the LMC does not display evidence for a characteristic scale in the star-formation process.  Similar results were found for the spiral galaxy NGC~628 (Elmegreen et al.~2006).

\subsection{Luminosity distribution}

\begin{figure}
\begin{centering}
\hspace{-0.5cm}
\includegraphics[width=8.5cm]{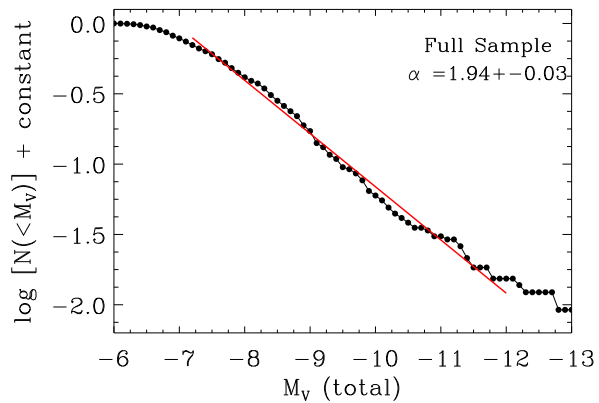}
\caption{The cumulative luminosity distribution of the total sample of groups.  The solid line shows the best fit to the data of the form $N(dL) \propto L^{-\alpha}dL$, where the best fitting $\alpha$ and error are given in the panel.} 
\label{fig:lumfunc}
\end{centering}
\end{figure}

The luminosity of each group was found by summing the V-band luminosity of all the stars found to be part of the group.  For simplicity, we just concentrate on the 'total sample' (i.e. all of the derived groups with multiple detections removed), although we note that the results for each breaking distance were similar.  The cumulative luminosity distribution is shown in Fig.~\ref{fig:lumfunc}, along with a power-law fit to the data of the form $NdL \propto L^{-\alpha}dL$, shown as a solid (red) line.  We find that the distribution is well fit by such an analytic form, with $\alpha=1.94\pm0.03$.  This is in excellent agreement with what was found for the group luminosity function in M33 (Bastian et al. 2007).  It is also in good agreement with the study of groups of OB stars in the SMC based on the MST method by Oey, King \& Parker~(2004). The implications of this are discussed in detail in Bastian et al.~(2007).  We note that the luminosity function may not directly relate to the mass function of the groups if there is a substantial age spread in the groups or if the initial mass function (IMF) of stars is dependent on environment (i.e. since we only sample the upper end of the stellar IMF we are insensitive to any fluctuations below our sample limits). 

\section{Spatial distributions of subsamples}
\label{sec:subsamples}

\subsection{Age groups and galactic structure}

The above sections dealt solely with young OB stars, however, using different colour and magnitude cuts 
in the CMDs of the LMC stars from the catalogue of Zaritsky et al.~(2004), we can select populations of different mean ages.  This is done by taking colour-magnitude regions (corrected for foreground extinction of 0.41, 0.32, 0.25, 0.15 mag for U, B, V, and I, respectively) along the main sequence and red-giant branches of the CMD.  In order to minimize the effect of larger contamination in the outskirts of the galaxy, and to assure complete geometrical coverage, we limited our selection to the inner three degrees (in radius) of the LMC.  

\begin{figure}
\hspace{-0.5cm}
\includegraphics[width=8cm]{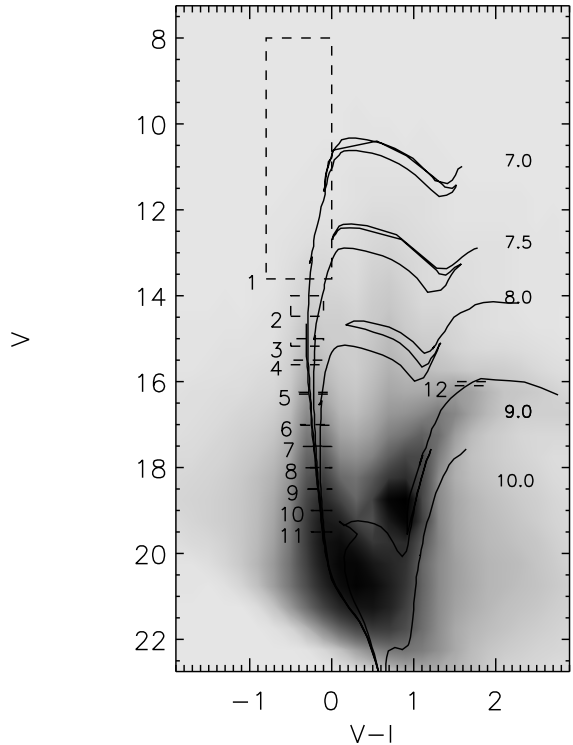}
\caption{Colour-magnitude diagram used to select the boxes.  The size of the boxes was determined by requiring 2000 sources in each colour-magnitude box.  Stellar isochrones from the Padova group (Girardi et al.~2002) are shown for logarithmic stellar ages (in years) of 7, 7.5, 8, 9, \& 10 and Z=0.08 metallicity.} 
\label{fig:cmd}
\end{figure}

\begin{figure}
\includegraphics[width=8.5cm]{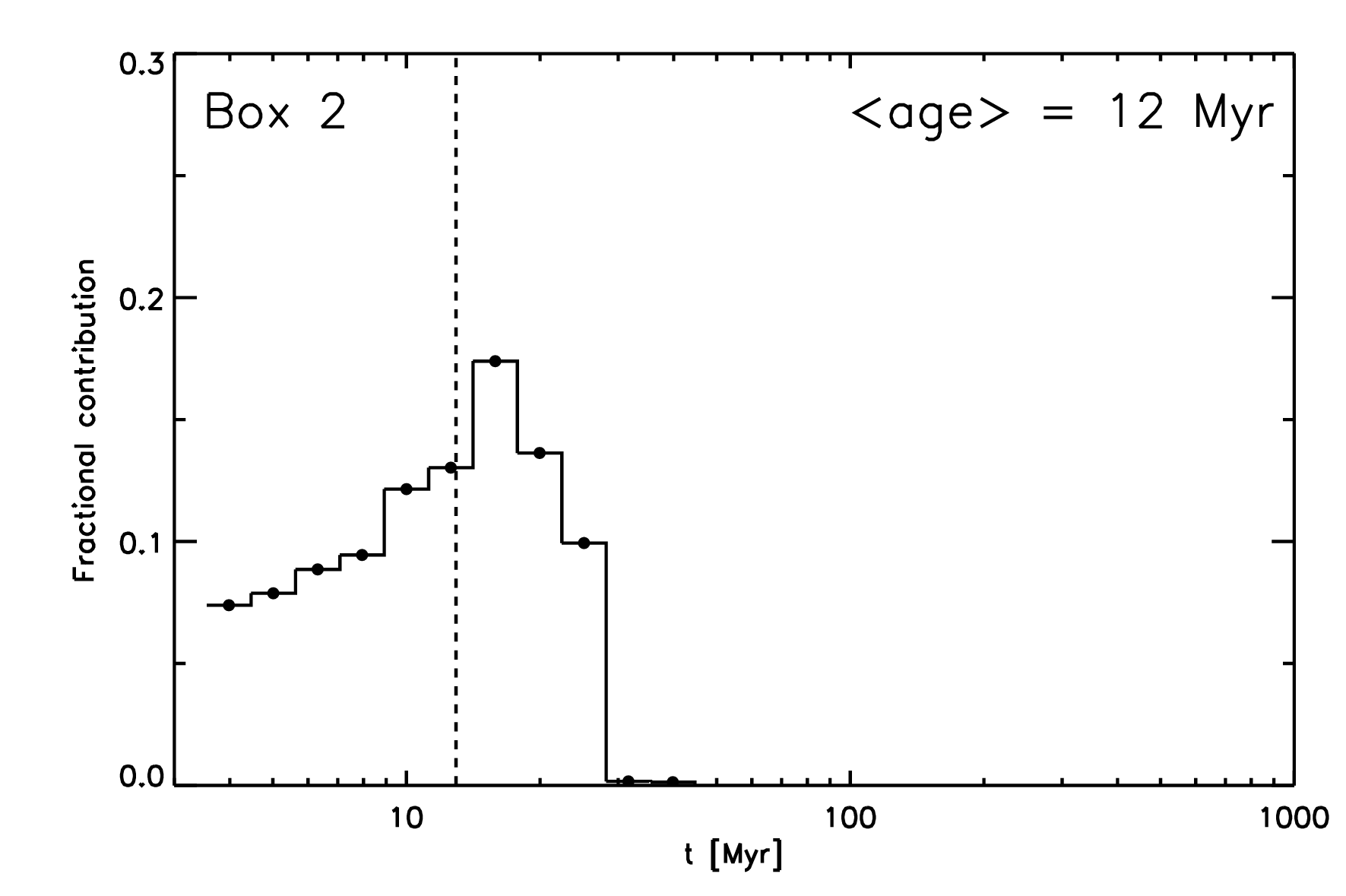}
\includegraphics[width=8.5cm]{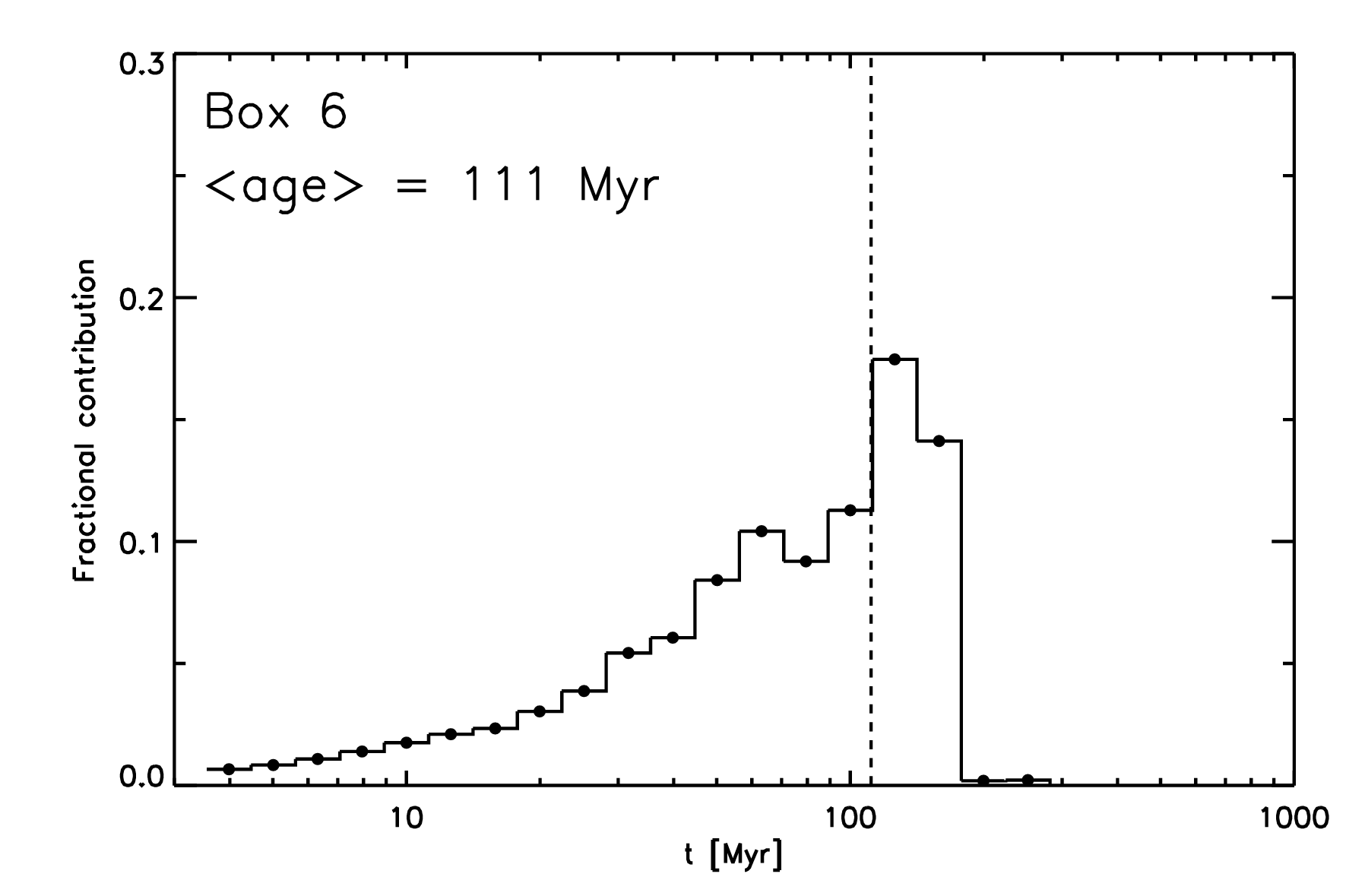}
\caption{Examples of the age dating process.  The histogram shows the fraction of stars which fall in our colour-magnitude selection box. The dashed line shows the mean age of all the stars which fall in the colour-magnitude box.   See text for details.} 
\label{fig:age-windows}
\end{figure}

For selection of the age boxes we follow the same procedure as Gieles, Bastian, \& Ercolano~(2008, hereafter GBE08).  The adopted regions are shown in Figure~\ref{fig:cmd}, where the sizes of the boxes were chosen such that they enclose 2000 stars within their boundaries.  In order to determine the average age of each of the boxes, we constructed a synthetic population of stars, assuming a constant star-formation rate and a random sampling from a Salpeter initial mass function (1955).  We used the Padova isochrones (Girardi et al. 2002 and references therein, Z=0.008) to assign a magnitude and colour for each model star, and assume a distance to the star of 50~kpc.  We then find the number of  stars at each age which fall into our colour-magnitude regions, and take the mean age of the stars which are contained within the designated region.  Two examples (boxes 2 and 6) are shown in Fig.~\ref{fig:age-windows}.   The selection cuts and mean ages are shown in Table~\ref{table:cuts}.  We note that using the median age of each box instead of the mean does not significantly change our conclusions.  Since we are using a CMD to select stars, the primary parameter that we are selecting is in fact stellar mass and not stellar age.  We implicitly assume that high and low mass stars (above a few solar masses) have the same spatial distribution.

\begin{table}
\begin{center}
{\scriptsize
\parbox[b]{8cm}{
\centering
\caption[]{Colour/magnitude cuts and corresponding ages for different 'boxes' in the LMC (shown graphically in Fig.~\ref{fig:cmd}).}
\begin{tabular}{c c c c }
\hline
\noalign{\smallskip}
Box   &  $V-I$ & $V$ &  mean age \\ 
                     &  (mag) & (mag)    &    (Myr) \\ 
\hline
1 & -0.8 -- 0.0 & 8.0 -- 13.61 & 9 \\ 
2 & -0.5 -- -0.1 & 14.0 -- 14.48 & 13 \\ 
3  & -0.5 -- -0.1 & 15.0 -- 15.17 & 31 \\ 
4  & -0.5 -- -0.1 & 15.5 -- 15.61 & 43 \\ 
5 & -0.4 -- 0.0 & 16.25 -- 16.29 & 54 \\ 
6 & -0.38 -- 0.0 & 17.0 -- 17.02 & 111 \\ 
7 & -0.35 - 0.0 & 17.5 -- 17.51 & 128 \\ 
8 & -0.32 -- 0.0 & 18.0 -- 18.01 & 176 \\ 
9 & -0.30-- 0.0& 18.5 -- 18.50 & 195 \\ 
10 & -0.25 -- 0.0 & 19.0 -- 19.00 & 219 \\ 
11 & -0.25 -- 0.0 & 19.5 -- 19.50 & 248 \\ 
12 & 1.5 -- 1.9 & 16.0 -- 16.10 & 1008 \\ 

\noalign{\smallskip}
\noalign{\smallskip}
\hline
\end{tabular}
\label{table:cuts}
}
}
\end{center}
\end{table}

\begin{figure*}
\hspace{-0.5cm}
\includegraphics[width=18cm]{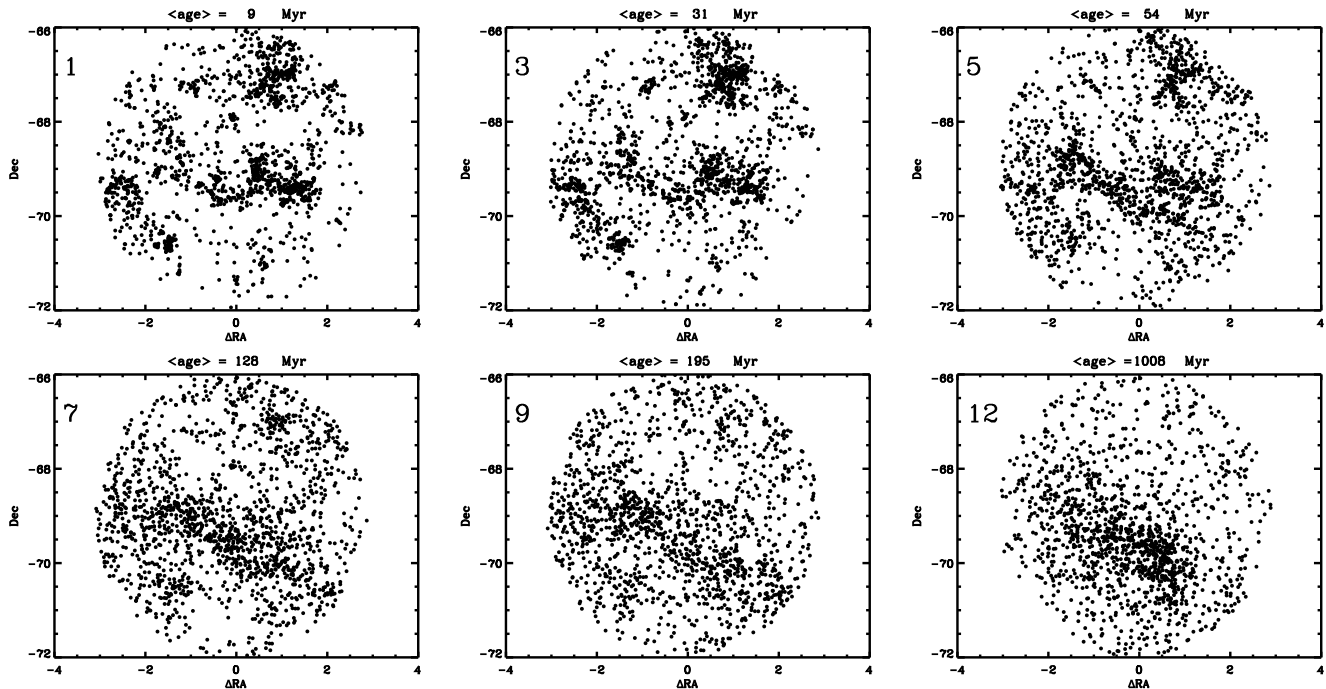}
\caption{The spatial positions of the stars in boxes 1, 3, 5, 7, 9, \& 12.  Only the inner three degrees in radius are used in the analysis.  The bar of the LMC is clearly visable in older populations, and dominates in age box 12.} 
\label{fig:spatial}
\end{figure*}

Once the boxes have been selected we can then look at the spatial positions of the stars for each box.  Six examples  are shown in Fig.~\ref{fig:spatial} where the box number and mean age are given in each panel.  It is clear that the younger populations contain considerably more substructure than the older populations.  The LMC bar is clearly visable in the older boxes, namely boxes 7, 9, and 12 in Fig.~\ref{fig:spatial}, and is clearly dominating the spatial distribution in box 12.  In the next section we will quantify these differences.

\section{Spatial analysis - quantifying structures} 
\label{sec:spatial}
\pagebreak[4]

\subsection{Q parameter}
\label{sec:q}

Cartwright \& Whitworth~(2004) have developed a formalism to quantify the structure of young star clusters.  Their method uses the normalised mean MST branch length, \mbar, and the normalised distance between all sources within a region, \sbar.  The ratio between these two quantities, \mbar/\sbar = $Q$, is able to distinguish between a power-law (centrally concentrated) profile and a profile with sub-structure (i.e. a fractal distribution).  Additionally, this parameter can quantify the index of the power-law or the degree of sub-substructure, which, if one assumes is due to a fractal nature, its fractal dimension can also be estimated.  Assuming a three dimensional structure, if $Q$ is less than 0.79 then the region is fractal, larger than 0.79 refers to a power-law structure, and a value of 0.79 implies a random distribution (a 3D fractal of dimension 3 is a random distribution).  However, if the distribution is two dimensional, as is approximately true if one is looking at a disk-like galaxy face on (like the LMC), then a random distribution has a $Q$ value of 0.72.

While this method was developed and has been applied to young embedded clusters (Cartwright \& Whitworth~2004, Schmeja \& Klessen~2006, Kumar \& Schmeja~2007), it has also recently been applied to full stellar populations within the SMC (GBE08), finding that the Q parameter increases with the mean stellar population age. 

During our analysis, we found an unexpected relation between the measured $Q$ value and the elongation parameter introduced by Schmeja \& Klessen~(2006), in that more elongated clusters have a lower $Q$-value.  We note here, that while this does not affect our results (each box has an elongation less than 1.2) it may be a serious problem for young embedded clusters which often have a filamentary structure (e.g. Gutermuth et al.~2008).  This issue and a proposed solution are dealt with in more detail in Appendix~A.  Additionally, in Appendix~B we discuss the effects of differential extinction on the $Q$ parameter.

In the bottom panel of Fig.~\ref{fig:panel} we show the measured $Q$-value for each of our 12 boxes (filled circles).   The $Q$-value is clearly increasing with age, beginning at a value of $Q=0.57$.  In order to measure the age at which the spatial distribution becomes uniform, we fit the $Q$-value as a function of the logarithm of age (shown as a dashed line in the bottom panel of Fig.~\ref{fig:panel} and calculate when this function crosses that of a flat distribution (i.e. when the $Q$-value becomes constant).  This happens at $\sim168\pm30$~Myr, where the error was calculated by the uncertainty in the value of the flat region of the diagram, with a value of 0.73.  

This is similar to what is expected for a 2D slightly centrally concentrated stellar distribution (a 2D power-law distribution with index $0.3$, decreasing outward; Cartwright \& Whitworth 2004), as is our reference distribution (see \S~\ref{sec:tpcf}).  This indicates that the spatial distribution of the stars becomes more homogeneous as the populations age.  However, box 12 has a much larger $Q$-value, which we attribute to the LMC bar, a strongly centrally concentrated 3D structure in the galaxy.  Errors were estimated by generating artificial populations with the same number of sources, deriving their $Q$-value, and taking the standard deviation.

\subsection{Two point correlation function}
\label{sec:tpcf}

The second technique that we used to determine the timescale for structures to be erased in the LMC is the two-point correlation function (TPCF).  This method determines the distance between all possible pairs of stars (d$_{s}$), shown as a histogram, which is then normalized to that of a reference distribution, i.e. $N_{\rm links} ({\rm d}_{s})$ / $N_{\rm reference}({\rm d}_{s})$. Here,  $N_{\rm reference}$ is taken as taken to be a slightly centrally concentrated powerlaw with index 0.3, which is a good representation of the inner three degrees of the LMC for stars with V $<$ 20.   This is similar to that done by Gomez et al.~(1993) who used the TPCF to study the distribution of pre-main sequence stars in Taurus,  although these authors used a random distribution as their reference distribution.  We are then left with a histogram for each age box of the number of connections with a certain distance, relative to the reference distribution.  For each of the age boxes we fit the distribution (in a logarithm of distance  vs. logarithm of the number) with a linear relation to get the slope and zero-point.

\begin{figure}
\includegraphics[width=8.5cm]{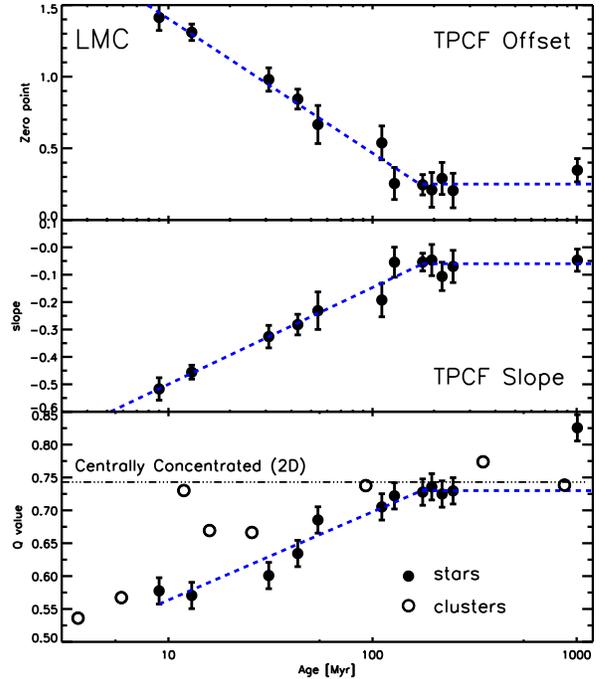}
\caption{The results of the three measurements (two techniques) of the erasure of substructure within the LMC.  The top and middle panels show the measured zero points and slopes of the two point correlation function for all age boxes.  As the reference distribution we use a centrally concentrated power-law distribution with index 0.3.  Note the smooth transition from younger (more substructure) boxes to older ones.  Both the zero point and slope distributions become flat at $\sim175$~Myr, suggesting that most substructure has been erased by this age.  The results of the $Q$-parameter are shown in the bottom panel, which largely agree with the results of the TPCF, i.e. that structure is erased in $\sim170$~Myr.  Additionally, we show the evolution of $Q$ for the cluster sample of Hunter et al.~(2003).} 
\label{fig:panel}
\end{figure}

If the distributions are becoming more similar to the reference distribution, then the slope and zero-points should tend towards values of 0 (see Appendix~\ref{appendix:contamination} for a demonstration of this).  The results for all age boxes are shown in the top and middle panels of  Fig.~\ref{fig:panel}.  Similar to the $Q$-parameter method, we find that the distribution for the boxes is approaching the reference distribution as the mean stellar age increases, for both the zero-point and slope of the distributions.  The fact that the values do not go precisely to zero is most plausibly due to a slight mis-characterization of the reference distribution.  Errors on each of the points were calculated by randomly selecting 1100 stars in the age box and repeating the analysis.

As can be seen by comparing all three panels in Fig.~\ref{fig:panel}, the three measurements of the disappearance of substructure agree very well.  For the two measurements provided by the TPCF, we fit the zero-point and slope as a function of the logarithm of age (shown as dashed lines), and find that this crosses the flat portion of the diagrams at $175\pm25$~Myr, where the error indicates the uncertainty in defining the flat region of the diagram.

\subsection{Comparison  of the methods}

The three measurements\footnote{While we have three measurements, we caution that only two are independent, as the slope and zero-point of the TPCF are intimately related.} of the erasure of substructure within the galaxy (the $Q$-parameter, and the slope and zero-point of the TPCF) are all in close agreement.  

{\it These results show that stars are formed with a large amount of substructure in the LMC and that, as this population evolves, it becomes more uniform, becoming statistically indistinguishable from the background reference distribution within $\sim175$~Myr.}  Through the $Q$-parameter, and comparing with artificially generated distributions, as was done in Cartwright \& Whitworth~(2004), we estimate that stars are born with a fractal distribution, with a 2D fractal dimension of $\sim1.8$.

Such evolution of stellar structures has also been seen in NGC 1313 (Pellerin et al. ~2007) and NGC 4449 (Annibali et al. 2007) using HST-ACS data.  Both studies found that massive stars (OB stars) contain more structure than older stellar populations.  Pellerin et al. (2007) explain this evolution of structure as due to 'infant mortality', i.e. that young stars are often formed in dense clusters which then 'pop' (Kroupa~2002) when the residual gas left over from the star-formation process leaves the cluster (e.g. Bastian \& Goodwin 2006).  Due to our ability to quantify the timescale and degree of clustering we can rule out this scenario, which we will return to in \S~\ref{sec:interpretations}.

In such statistical analyses as presented here, contaminating sources may potentially seriously affect the results.  In order to estimate the magnitude of these effects we have run a series of simulations where we generate artificial distributions with known parameters, and substitute random (or reference) distributions with a given percentage.  The results of these experiments are shown and discussed in Appendix~\ref{appendix:contamination}.  In particular, we note that young stars will contribute to all age boxes (with the exception of box 12), and since they display a larger degree of substructure, some degree of substructure is expected to be present in all boxes.  However, the contamination is expected to be low (much less than 30\%), hence the statistical methods should not be affected by them, although they can be seen by eye in Fig.~\ref{fig:spatial}.

\section{Comparison of optical groups, young stellar objects and stellar clusters}
\label{sec:comparison}

In order to further understand the evolution of stellar structures within the LMC we take advantage of two recent surveys.   The first is that of Whitney et al.~(2008) who used the SAGE (Meixner et al.~2006) multi-band Spitzer survey of the LMC to select high probability young stellar objects (YSOs).  These populations are expected to trace the evolution of forming stars/clusters from the parent clouds, through the embedded phase to the point where they can be detected at optical wavelengths.  The second survey was that of Hunter et al.~(2003) who catalogued the cluster population of the LMC, and measured spatial positions, magnitudes, and ages for over 900 clusters.

Constraining the YSO sample to the inner three degrees of the LMC (as was done for the optically selected samples) we find $Q_{\rm YSO} = 0.68$.  This is significantly higher than that found for the young optically selected (OB star) sample ($Q_{\rm OB}=0.57$).  Since AGB stars and YSOs have similar colours even in the mid-IR (e.g. Whitney et al.~2008) it is possible that this difference is due to contamination of the YSO sample.  If it is due to contamination (see Appendix~\ref{appendix:contamination}) then this implies that the YSO sample has approximately $40\%$ contamination from stars which have a similar distribution to that of our reference distribution.  This may be the case if AGB stars are the main contamination.  The alternative interpretation, is that the YSO sample is made up largely of lower-mass stars than our OB star sample (Whitney et al.~2008), and that the distribution of star formation is mass dependent.  Such extreme spatial dependence on mass appears to be at odds with observations of young Galactic star-forming regions (as indicated by the similarity of the stellar IMF in different environments).  Finally, we note that the lower spatial resolution of the YSO sample may act as to raise the measured $Q$-value.

For the cluster distribution, we again limit the sample to the inner three degrees in order to draw useful comparisons with our optically selected samples, and additionally provide a magnitude cut in the data of $M_{V} < -4.0$ in order to avoid incompleteness effects in the central region.  We then break the cluster sample up into 8 age bins, with each bin containing an equal number of clusters, namely 69.  The resulting distribution of $Q$-parameters vs. cluster age is shown in the bottom panel of Fig.~\ref{fig:panel}.  As seen in the stellar distributions, the $Q$-parameter of the cluster populations increases with mean age, and becomes flat between $100-200$~Myr.  The larger scatter in the cluster points relative to the stellar boxes is due primarily to the lower number of points used to calculate $Q$ ($69$ vs. $2000$).  Overall, the cluster population loses structure in a similar way to the stellar population, and if anything, the structure removal is faster for the clusters.  We will return to this point in \S~\ref{sec:interpretations}.

\subsection{Interpretations: Infant mortality vs. general galactic dynamics}
\label{sec:interpretations}

\subsubsection{Infant Mortality}
\label{sec:im}

It has been shown that the majority of clusters that form become disrupted within the first few 10s of Myr of their lives (e.g. Lada \& Lada 2003).  The mechanism believed to be responsible for this is the rapid removal of gas, due to stellar winds and/or supernovae, from the natal cluster.  This gas removal leaves the cluster in an unbound state due to the removal of a large fraction of the gravitational potential of the cluster, that leaves the remaining stars in a super-virial state.  If the star-formation efficiency is low ($<30\%$) the entire cluster may become unbound and expand rapidly.  These types of clusters have been referred to as 'popping clusters' (Kroupa~2002) and evidence of their expanding nature has been seen in their low light-to-mass ratios for a given age (Goodwin \& Bastian 2006).  The clusters are predicted to expand with a velocity close to their velocity dispersion and  may have a severe effect on the observed galactic structure.

The standard scenario for the 'popping' clusters is that the stars and gas are in, or close to, virial equilibrium when the natal gas is removed.  Thus, we can use the virial equation to find the relation between a cluster's mass (gas+stars) and velocity dispersion (e.g. Spitzer 1987).  Due to the relatively low star-formation rate in the LMC we do not expect a large population of massive clusters.  Assuming a stochastically sampled power-law mass function of the clusters (gas+stars) with index $-2$, lower mass limits of $M_{\it low} = 1, 10, \&~ 100~M_{\odot}$, and cluster radii of 1~pc, we can set up a simple model of the evolution of the structure of a galaxy due to infant mortality.  

We start with a spatial distribution of clusters with the same $Q$-value as the youngest observed clusters.  At each of these cluster centres we place a cluster of mass \mcl~ and radius 1~pc  We then assign a number of stars to each cluster, directly proportional to its mass, with a gaussian spatial distribution (stochastically sampled).  The velocity dispersion of each cluster (gas+stars) is derived through the virial equation, assuming a given star-formation efficiency (SFE, c.f. Goodwin \& Bastian~2006).  The cluster expansion is mimicked by increasing \rcl, with \rcl$= t_{\rm dis}$\vcl, where $t_{\rm dis}$ is the time since the cluster began expanding and \vcl~ is the cluster velocity dispersion. At each timestep, we select 1000 stars at random and calculate their $Q$-value.  Only stars within the inner 3 degrees (the simulations were set up relative to the observed LMC distribution) were selected.

The results of these simulations are shown in Fig.~\ref{fig:evo}, where the top, middle and bottom panels represent simulations where the mass function of the clusters is sampled down to 100, 10, \& ~1~M$_{\odot}$ respectively.  Since we have used the same number of clusters in each simulation, the horizontal shift in the simulations is related to the average velocity dispersion of the clusters, which decreases with decreasing M$_{\rm low}$.  If clusters start more compact (as suggested by observations of young clusters, e.g. Mackey et al.~2007, Bastian et al.~2008), this would shift the curves to the left.

The cluster mass function is thought to be continuous down to single stars (e.g. Oey et al.~2004), arguing for the applicability of the simulations with M$_{\rm low}$ = 1~M$_{\odot}$.  This suggests that the erasure of structures is not being driven by 'popping clusters'.  However, a caveat to these simulations is that we have assumed that all clusters pop in the same manner, regardless of mass.  This, however, may not be true, as the number of O-stars (which drive the gas removal, and hence the 'popping') in the cluster may influence the expansion properties (e.g. Baumgardt \& Kroupa~2008).  Additionally, we have not taken the motion of the clusters through the galaxy into account which also destroys substructure (see \S~\ref{sec:galdyn}).

We can test the effect of infant mortality another way, namely by looking at the evolution of the $Q$-value for stars and that for catalogued star clusters.  If infant mortality was a dominant cause of structural evolution within the galaxy, then we would expect the $Q$-values of the stars to increase more rapidly with age than that of clusters.  This is because the star clusters which survive the 'infant mortality' phase will not have their positions changed, whereas the stars that formed in clusters that 'popped' will be expanding away from their natal positions. In other words, the stars will have two components to their velocity, namely expansion and motions through the galaxy, whereas clusters will only have one component, the velocity dispersion of the galaxy.

For the cluster catalogue we use the Hunter et al.~(2003) survey presented in \S~\ref{sec:comparison}.  In the bottom panel (open circles) of Fig.~\ref{fig:panel} we show the measured $Q$-value for each of the cluster sub-samples.  In general we see the same behavior as found for the stellar boxes, namely that the $Q$ value increases with mean age, beginning with a highly fractal, or substructured distribution and becoming flat near the same value as that found for the stellar boxes.  The evolution between the stars and clusters is quite similar, in fact the clusters seem to evolve slightly faster (although this does not appear to be statistically significant given the small number of clusters per age bin) than the stars.  Hence we conclude that the same mechanism is erasing the substructure seen in the stellar and cluster distributions, which eliminates the possibility that 'popping clusters' are the dominant cause, although this effect is likely taking place within the galaxy.


\subsubsection{Galactic Dynamics}
\label{sec:galdyn}

We conclude that the dominating effect of structure evolution in the LMC is general galactic dynamics.   It is the velocity dispersion of the galaxy which eliminates substructures and pushes the stellar distributions toward a random, or background, distribution.  

These results are consistent with the findings of the distribution of Cepheids in the Galaxy (Efremov 1995) and the LMC (Marconi et al.~2006).  For the Galaxy, Efremov (1995) found evidence for the presence of large 'Cepheid clouds' with sizes of a few hundred parsecs, which, considering the age of Cepheids (a few 10s to 100s of Myr) is what is expected from our results.  More directly, Marconi et al.~(2004, their Figs.~4~\&~5) found that young Cepheids ($20-80$~Myr) display much more substructure in their spatial distribution than older Cepheids ($>80$~Myr) whose distribution is becoming closer to random, in agreement to that found by Elmegreen \& Efremov~(1996).  Finally, we note that this is a similar timescale ($\sim150$~Myr) to that found by Efremov \& Elmegreen~(1998) for the average separation between pairs of star clusters in the LMC to become constant, i.e. for the cluster population to become statistically uniform.

The LMC has a velocity dispersion of $\sim22$~km/s (Carrera et al.~2008), if we adopt a radius of the LMC of $3-4$~kpc this translates to a crossing time ($t_{\rm cross} = R/\sigma$) of $\sim135-180$~Myr.  This is remarkably similar to the timescale derived for the elimination of structure within the LMC.  From this, we conjecture that structure will be erased on a crossing-time, independent of spatial scale.  This may explain why young lower-mass clusters often contain large amounts of substructure (e.g. Gutermuth et al.~2005), while many of the massive young clusters (such as Orion) have much smoother central regions, where the crossing time is significantly lower than the outer regions (see Elmegreen~2008).

\begin{figure}
\includegraphics[width=8.5cm]{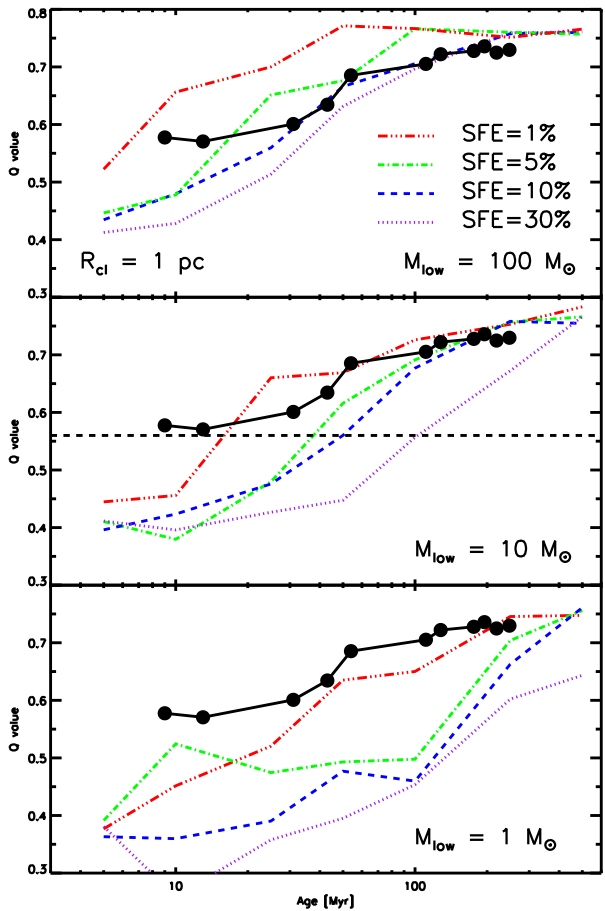}
\caption{Simulated evolution of a galaxy due to infant mortality for various star formation efficiencies (SFE), see text for details of the simulations.  Three values of the lower mass limit of the cluster mass function are given.  All simulations begin with clusters with radii of 1~pc.  The observed change in Q vs. time is shown as filled circles. } 
\label{fig:evo}
\end{figure}

In GBE08 we show the same quantitative relationship between the stars and clusters in the SMC and reach a similar conclusion, namely that popping clusters are not driving the evolution of substructure within the galaxy.  In the SMC, substructure is removed in $\sim80$~Myr, which is again similar to the galactic crossing time.  Note that the erasure of structure in the 'popping cluster' scenario (e.g. Fig.~\ref{fig:evo}) is mainly dependent on the average internal velocity dispersion $\sigma_{\rm mean}$ of the clusters.  Since $\sigma_{\rm mean}$ is independent of the number of clusters in a population (it only depends on the mass function index and lower mass limit of clusters - which we assume to be invariant) then all galaxies should have structure erases in approximately the same time.  However, in the timescale of structure erasure is twice as fast as in the LMC.

An interesting difference between the SMC and LMC stellar and cluster distributions is the $Q$-values which they attain when the become flat.  For the SMC the $Q$-value of both clusters and stars becomes flat at $Q=0.81$ whereas for the LMC, both distributions become flat at $Q=0.73$.  This difference shows the ability of the method to quantify the structure of distributions, as the SMC is largely three dimensional whereas the LMC is mostly disk-like, or two dimensional (both slightly centrally concentrated).  The expected $Q$-values for homogeneous 2D and 3D distributions are 0.71 and 0.79, respectively (Cartwright \& Whitworth~2004), therefore explaining the observed difference between the $Q$-values in the LMC and SMC.

\section{Conclusions}\label{sec:conclusions}

We have presented an analysis of the spatial distribution of stellar structures within the LMC.  By using the Zaritsky et al.~(2004) Magellanic Cloud Photometric Survey and sampling different parts of the stellar colour-magnitude diagram, we have selected stellar populations of varying mean age.  This is supplemented with other datasets, namely the Whitney et al.~(2008) Young Stellar Object (YSO) infrared sample and the star cluster catalogue of Hunter et al.~(2003).

Based on the analysis of the distribution of OB stars, we confirm our earlier results on M33, namely that the star-forming groups are well approximated by a power-law luminosity function with index of $-2$ and  that there is no preferred size in the distribution of star-forming sites (e.g. that of the OB association).  Instead, we find a continuous distribution of sizes, and find that previous reports of a characteristic size were due to resolution effects, along with choosing, apriori, a single size-scale to probe.

We have used two methods to quantify the amount of substructure within a galaxy, the Q method (Carthwright \& Whitworth 2004), and the two-point correlation function (TPCF).  
We found that stars are formed in a highly structured way (approximating the distribution as a 2D fractal we find the fractal dimension to be $\sim1.8$) and that this distribution evolves rapidly, becoming indistinguishable from the background or reference distribution after $\sim175\pm30$~Myr.   We also found that the distribution of stellar clusters follows that of the stars of similar age, adding confidence in our age dating method as well as our structural analysis.


Thus, our findings suggest that stars are formed in a highly substructured distribution and that this distribution evolves rapidly, with small scale structures dispersing first, and that all of the original structure is removed in  $\sim175$~Myr.   This is approximately the crossing time of the galaxy and as such, we suggest that structures in galaxies, presumably regardless of spatial scale, are eliminated in a crossing time.  Similar analysis done on the SMC (GBE08) finds that substructure is erased on $\sim75$~Myr, which is comparable to the crossing time of that galaxy.  

By comparing the observations to simple models, we conclude that infant mortality (or 'popping star clusters') have a negligible effect on the stellar distribution in the LMC and that general galactic dynamics is the major driver of structural change.  
 It is worth noting that our results are not necessarily in contradiction with the 'infant mortality' scenario, but they show that this mechanism does not explain the observed evolution of stellar structure in galaxies.

Only a small fraction of stars are born in stellar clusters which will survive long enough to be able to be identified as 'clusters' in optical surveys (e.g. Lada \& Lada~2003, Gieles \& Bastian~2008, Bastian~2008).  Whether this is due to a high fraction of stars born in embedded 'clusters' which quickly disrupt or due to a large fraction of stars forming outside clusters (or more likely, some combination of these two extremes) is still an open question at the moment.

\section*{Acknowledgments}

We gratefully thank Simon Goodwin for discussions and help with generating fractal distributions.  We also thank Barbara Whitney for providing her list of YSO candidates before publication and Anil Seth for interesting discussions.  The referee, Jason Harris, is thanked for helpful comments and suggestions on the text, techniques, and results presented in this paper.  NB gratefully acknowledges the support of the Harvard-Smithsonian Center for Astrophysics, where a large part of this work was carried out.  BE was partially supported by {\it Chandra} grants GO6-7008X and GO6-7009X.

\appendix
\section{Effects of ellipticity on Q}
\label{sec:ell}
\begin{figure*}
\includegraphics[width=18cm]{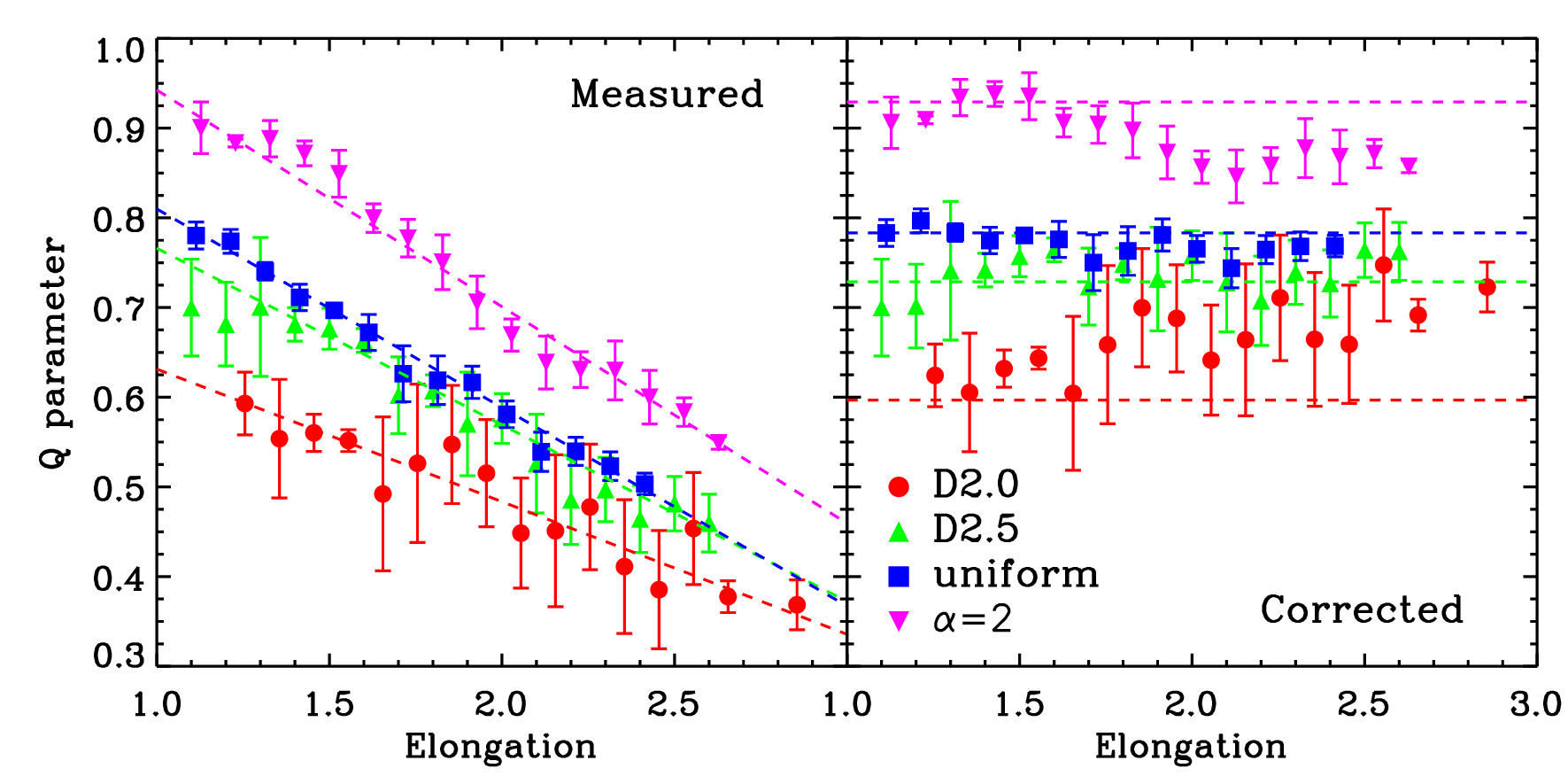}
\caption{The effect of elongation on Q.  {\bf Right:}  To correct for the effect of elongation we have taken the average slope of the simulations (shown in the left panel).  Doing this we find $Q_{\rm intrinsic} = Q_{\rm obs} - (-0.208*{\rm elongation}) - 0.229$.} 
\label{fig:q-elongation}
\end{figure*} 

In this section we investigate the influence of the elongation on the calculated Q parameter.  For consistency, we define the elongation in the same way as Schmeja \& Klessen~(2006), namely as the ratio of the derived radius when approximating the region as a circle to the radius when using the convex hull method.  Schmeja \& Klessen~(2006) give the relation between the elongation and the major to minor axis ratio of an ellipse.

In order to test the effect we first create 3D spherical models of varying fractal dimensions, uniform distribution and various centrally concentrated profiles, similar to what was done in Cartwright \& Whitworth~(2004).  We then multiply one of the three principle axes by a factor and measure the Q value and elongation along the three projected axes.  The results  for two fractal models (fractal dimensions of 2.0 and 2.5), a uniform distribution and a centrally concentrated distribution with power-law radial index of -2 are shown in Fig.~\ref{fig:q-elongation}.  It is clear that increasing the elongation decreases the measured Q value.  We have also fit each relation with a linear slope shown as a dashed line.  The error bars represent the standard deviation of three realisations of the simulations (the fractal models begin with a random seed).

While the slope of the change in Q as a function of the elongation is somewhat dependent on the starting Q value (i.e. what type of distribution is used) the overall trend is quite similar for all profiles considered.  As such, we use the average slope of the distributions in the left panel of Fig.~\ref{fig:q-elongation} in order to provide a correction function.  For this we find  $Q_{\rm intrinsic} = Q_{\rm obs} - (-0.208*{\rm elongation}) - 0.229$.  The right panel of Fig.~\ref{fig:q-elongation} shows the resulting values of the simulations (shown in the left panel) after this correction has been applied.  Here the dashed lines represent the input Q value (i.e. that determined  for the profile where no elongation has been applied).  While the given correction appears to correct well the simulations we note that large errors can be expected for systems with high elongations ($>1.5$, i.e. major/minor axis ratios greater than 2).

\section{Fractal models: extinction effects on Q}
\label{appendix:extinction}

An additional affect which may influence the measured $Q$ parameter is that of differential extinction.  This can be realized conceptually by imagining a uniform stellar population and putting a highly sub-structured distribution of gas/dust in the line of sight.  If the opacity is high enough, this will remove whole sections of the stellar distribution from sight, creating the appearance of a non-uniform stellar population.  Here, we we attempt to quantify this affect.
 
\begin{figure*}
\includegraphics[width=18cm]{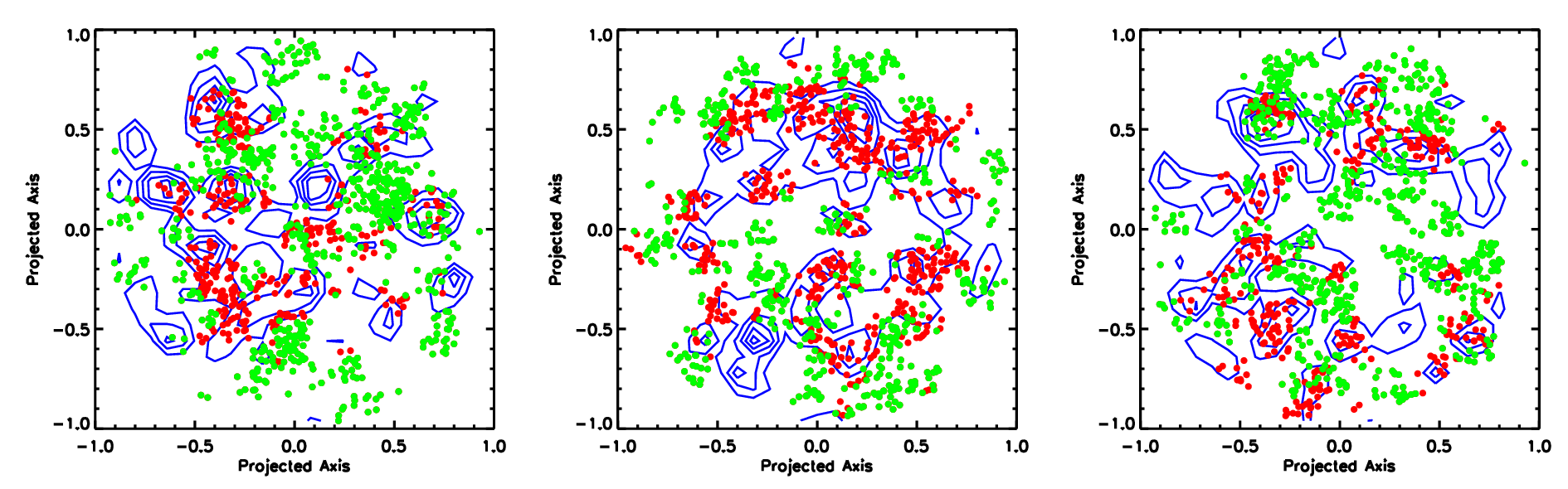}
\caption{Three projections of a simulation on the effect of extinction on the measured value of Q.   Both the gas particles and the stars are given a three dimensional fractal distribution of dimension 2.0.  The (blue) contours represent the number density distribution of gas particles in the simulations (integrated along the line of sight) while the green and red circles represent stars (i.e. together they make up the initial sample).  The red circles are stars which have been extincted and hence are not 'detected' while the green circles are stars which are not extincted enough to be removed from the sample (see text for details).  The full sample of stars has a Q value which is $0.07\pm0.03$ higher than that of the detected sample. } 
\label{fig:q-sims-figures}
\end{figure*}

For the simulations we make 3D fractal distributions of point sources with dimension 2.0 following the method presented in Goodwin and Whitworth (2004).  For each simulation we create 20,000 gas particles and 1,000 stellar particles independently.  We then chose a line of sight and counted the number of gas particles between the observer and each star within some projected tolerance, i.e. we find the surface density of gas between the observer and each star.  Stars which were found to have gas surface densities above a certain critical value were removed from the dataset.   Finally, we calculate Q for the original (non-extincted)  and extincted datasets and compare their values.  Three projections of one of the simulations are shown in Fig.~\ref{fig:q-sims-figures} where the contours represent the integrated column density of gas along the line of sight, red circles are stars which have been removed from the original dataset (i.e. are found to be extincted), and green circles represent stars which were not extincted (i.e. the red and green circles together make up the initial stellar distribution).

The main result is that extinction from clouds tends to lower the value of Q, where the change is related to the number of stars removed (i.e. the opacity of the clouds).  Typical changes in the derived Q value, $\Delta Q$, are lowered by 0.04 to 0.08 when 20\% or 50\% of the sources are extincted out of the sample, respectively.  This makes the distribution appear to have a lower fractal dimension than the actual distribution.  We have tested the effect of varying the number of particles used and the input fractal dimensions of the distribution and found that both effects are minimal.  Finally, we test the effect of culling the stars randomly and find that this does not change the value of Q ($\Delta Q<0.015$).

Since an unphysically large number of stars must be extincted out of our sample before Q is affected, we conclude that extinction does not significantly affect our results.

\section{Fractal models: effects of contamination}
\label{appendix:contamination}

Due to a variety of effects, some contamination may be expected to enter observational datasets.  This could be due to field stars in the Galactic halo or young (lower-mass) stars which have colours and magnitudes which fall in the CMD boxes used to selected older age stars.  In order to estimate the strength of these effects we have constructed the following experiments.  We start with a 2D projection of a 3D fractal model, chosen to have a similar Q-value as our youngest stellar age-box.  We then replace a fraction of the model stars, whose positions are chosen according to the probability function of a background distribution.  For the background we chose three distributions (all 2D distributions); a stochastic distribution, and two centrally concentrated distributions, both power-laws, with indices 0.3 and 1.0.  We used the prescription in Cartwright \& Whitworth~(2004) to construct the background distributions.  The centrally concentrated distribution with index 0.3 was chosen as this matches the projected radial profile of the LMC, when restricted to the inner three degrees in radius.

\begin{figure}
\includegraphics[width=9cm]{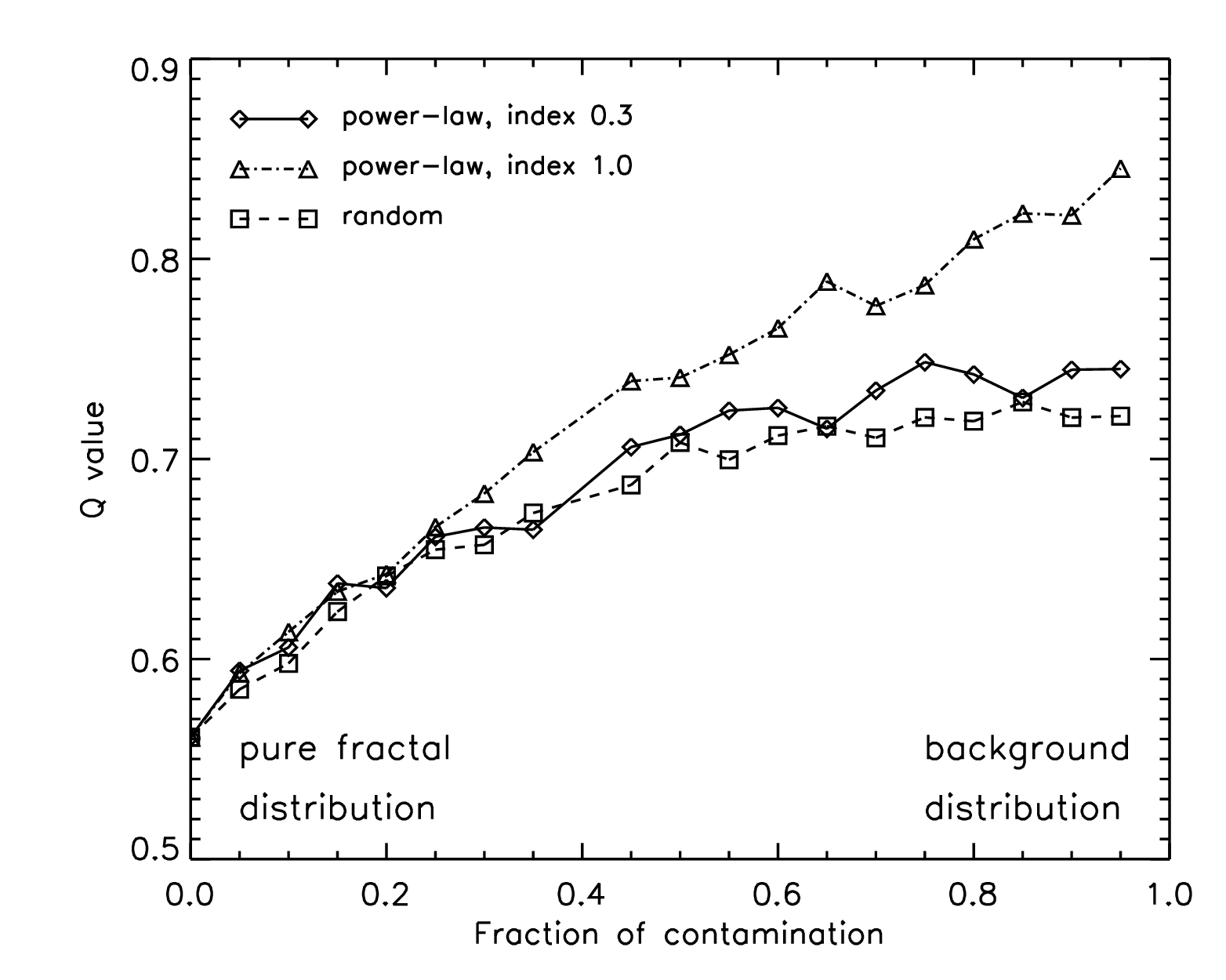}
\caption{The change in the Q-value of a fractal distribution when a given fraction of the stars are replaced by a background distribution.  Three examples are given for the background distribution, 1) a stochastic distribution, 2) a centrally concentrated power-law with index 0.3 (chosen to match the light profile of the LMC in the inner three degrees) and 3) a centrally concentrated power-law with index 1.0.} 
\label{fig:contamination-q}
\end{figure}

The results of the simulations are shown in Fig.~\ref{fig:contamination-q}.  In each case, we see the Q-value increase rapidly with increasing contamination.  In particular, the Q-value in the centrally concentrated case (with power-law index of 0.3) becomes largely flat for contamination fractions above 0.6.

This explains why small amounts of fractal substructure in the older age boxes (due to young stars which enter our selection box) do not significantly affect the results, as a large fraction of these stars would need to be present.

\begin{figure}
\includegraphics[width=9cm]{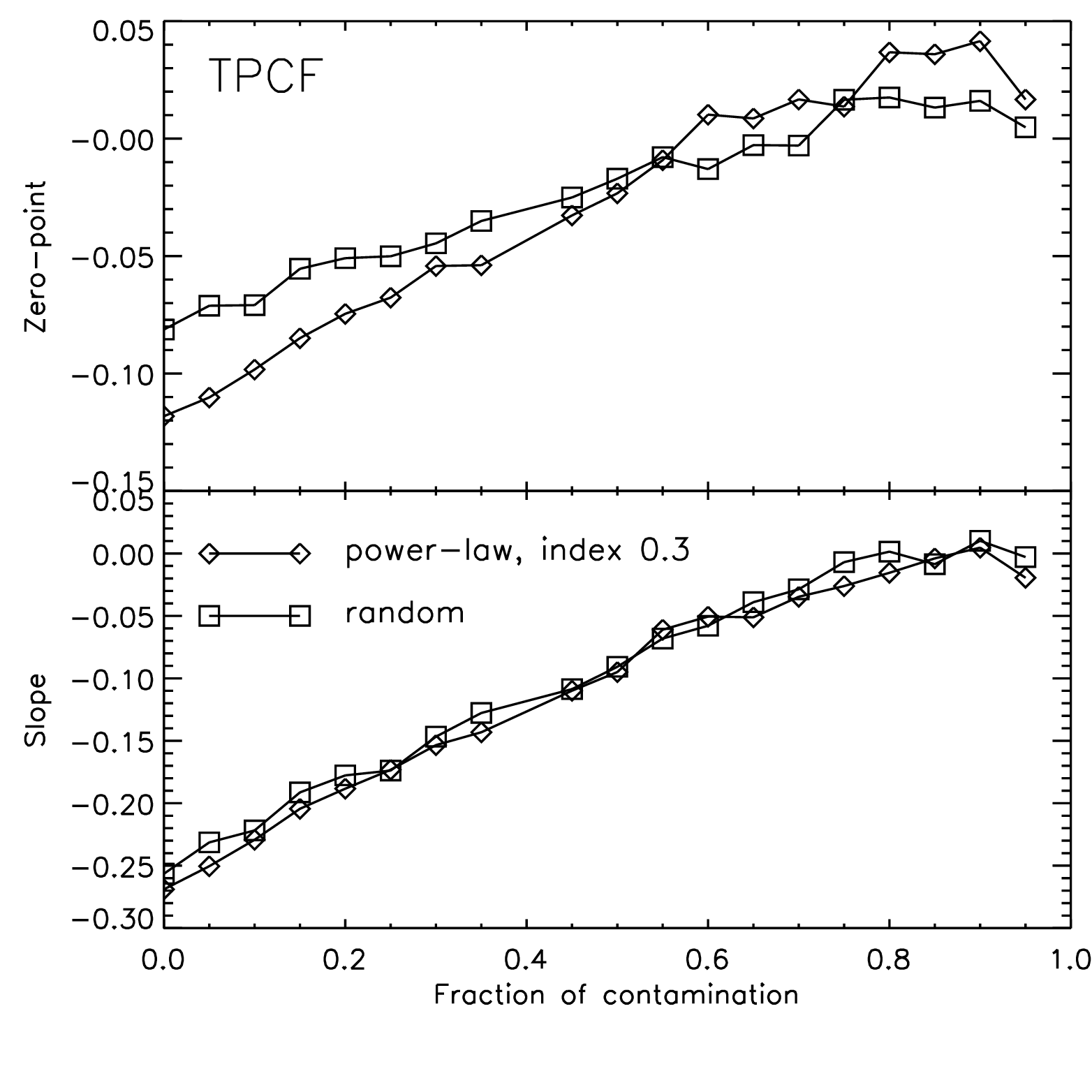}
\caption{The change in the slope and zero-point of the two-point correlation function  of a fractal distribution when a given fraction of the stars are replaced by a background distribution.  Two examples are given for the background distribution, 1) a stochastic distribution, 2) a centrally concentrated power-law with index 0.3 (chosen to match the light profile of the LMC in the inner three degrees).} 
\label{fig:contamination-tpcf}
\end{figure}

Figure~\ref{fig:contamination-tpcf} shows how the two-point correlation function analysis is affected by the same contamination.  As seen in the Q-distribution, the slope and zero-point of the TPCF smoothly transform from the value expected for a fractal distribution, to that expected for the background or reference distributions.  

Thus we conclude that while younger stars (which display more substructure) will contribute to all age boxes, this effect will not significantly alter our conclusions, as the expected contamination is low, their distribution will not significantly change the measured Q distribution or alter the TPCF.

\bsp
\label{lastpage}
\end{document}